%% file: 0-main.tex
\documentclass[]{spie}  

\usepackage{amsmath,amsfonts,amssymb}
\usepackage{graphicx}
\usepackage[colorlinks=true, allcolors=blue]{hyperref}

\title{Focal Plane Wavefront Sensing on SUBARU/SCExAO}

\author[a,b]{S. Vievard}
\author[c]{S.P. Bos}
\author[d]{F. Cassaing}
\author[a,e]{T. Currie}
\author[a,f]{V. Deo}
\author[a,b,g,h]{O. Guyon}
\author[i]{N. Jovanovic}
\author[c]{C.U. Keller}
\author[j]{M. Lamb}
\author[k]{C. Lopez}
\author[a]{J. Lozi}
\author[k]{F. Martinache}
\author[k]{D. Mary}
\author[c]{K. Miller}
\author[d]{A. Montmerle-Bonnefois}
\author[d]{L.M. Mugnier}
\author[k]{M. N'Diaye}
\author[l]{B. Norris}
\author[a]{A. Sahoo}
\author[d]{J-F Sauvage}
\author[a,f,m]{N. Skaf}
\author[c]{F. Snik}
\author[c]{M.J. Wilby}
\author[l]{A. Wong}

\affil[a]{National Astronomical Observatory of Japan, Subaru Telescope, 650 North Aohoku Place, Hilo, HI 96720, U.S.A.}
\affil[b]{Astrobiology Center of NINS, 2-21-1, Osawa, Mitaka, Tokyo, 181-8588, Japan}
\affil[c]{Leiden University, Rapenburg 70, 2311 EZ Leiden, Netherlands}
\affil[d]{ONERA, The French Aerospace Lab, University of Paris Saclay, F-92322 Châtillon - France}
\affil[e]{NASA-Ames Research Center, Moffett Field, California, USA}
\affil[f]{LESIA, Observatoire de Paris, Université PSL, CNRS, Sorbonne Universit\'e, Universit\'e de Paris, 5 place Jules Janssen, 92195 Meudon, France}
\affil[g]{College of Optical Sciences, University of Arizona, Tucson, AZ 85721, U.S.A.}
\affil[h]{Jet Propulsion Laboratory, 4800 Oak Grove Drive, MS 183-901, Pasadena, CA 91109, U.S.A.}
\affil[i]{California Institute of Technology, 1200 E California Blvd, Pasadena, CA 91125, U.S.A.}
\affil[j]{Dunlap Institute for Astronomy and Astrophysics, University of Toronto, 50 St. George St.,Toronto, ON, Canada}
\affil[k]{Université de la Côte d'Azur, Observatoire de la Côte d'Azur, CNRS, Laboratoire Lagrange, France}
\affil[l]{Sydney Astrophotonic Instrumentation Labs, University of Sydney, Australia}
\affil[m]{Department of Physics and Astronomy, University College London, London, United Kingdom}

\authorinfo{Further author information: Sebastien Vievard: E-mail: vievard@naoj.org}
\begin{document} 
\maketitle

\begin{abstract}
Focal plane wavefront sensing is an elegant solution for wavefront sensing since near-focal images of any source taken by a detector show distortions in the presence of aberrations. Non-Common Path Aberrations and the Low Wind Effect both have the ability to limit the achievable contrast of the finest coronagraphs coupled with the best extreme adaptive optics systems. To correct for these aberrations, the Subaru Coronagraphic Extreme Adaptive Optics instrument hosts many focal plane wavefront sensors using detectors as close to the science detector as possible. We present seven of them and compare their implementation and efficiency on SCExAO. This work will be critical for wavefront sensing on next generation of extremely large telescopes that might present similar limitations.
\end{abstract}


\input{1-Introduction}

\input{2-ZAP}


\input{4-LAPD}

\input{5-MonoPD}


\input{6-FandF}


\input{7-NN}

\input{8-DRWHO}

\input{9-Conclusion}

\section{Acknowledgments}
The development of SCExAO was supported by the Japan Society for the Promotion of Science (Grant-in-Aid for Research \#23340051, \#26220704, \#23103002, \#19H00703 \& \#19H00695), the Astrobiology Center of the National Institutes of Natural Sciences, Japan, the Mt Cuba Foundation and the director's contingency fund at Subaru Telescope. F. Martinache's work is supported by the ERC award CoG - 683029. S.V. would also like to thank Julien Milli for the discussions on the Low Wind Effect. The authors wish to recognize and acknowledge the very significant cultural role and reverence that the summit of Maunakea has always had within the Hawaiian community. We are most fortunate to have the opportunity to conduct observations from this mountain.

\bibliography{report} 
\bibliographystyle{spiebib} 

\end{document}

%% file: 1-Introduction.tex
\section{INTRODUCTION}
\label{sec:intro}  

Imaging and characterizing circumstellar environments is the main science goal for many high contrast imagers installed on large ground-based telescopes. The combination of Adaptive Optics (AO) - and even extreme AO (XAO) - systems and starlight suppression techniques have allowed, during the last 12 years, for direct imaging of many massive planets~\cite{Marois2008,Lagrange2010,Currie2014} or protoplanets~\cite{Keppler2018,Haffert2019,Currie2015} orbiting nearby young stars. With the construction of Extremely Large Telescopes~\cite{Sanders2013,gilmozzi2007european} (ELTs) on the horizon, the hope is to enable direct imaging and spectroscopic analysis of potentially habitable planets, or the unravel of the earliest stages of planet formation.

\begin{figure}[!h]
    \centering
    \includegraphics[width=0.7\linewidth]{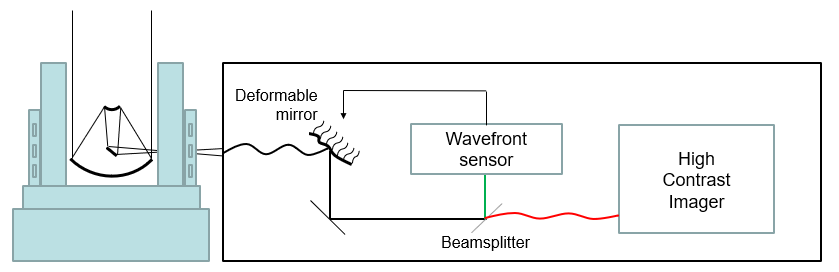}  
    \caption{Basic principle of an AO system. The light coming from the telescope is split between a sensing path (green) and a science path (red). In the sensing path, the WFS measures the wavefront perturbations and corrects them by changing the shape of the deformable mirror. The differential aberrations in the science path are not seen by the WFS.}
    \label{fig-NCPA}
\end{figure}{}

\begin{figure}[!h]
    \centering
    \includegraphics[width=0.4\linewidth]{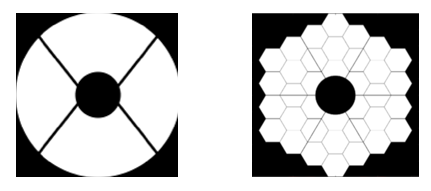}  \\ \includegraphics[width=0.6\linewidth]{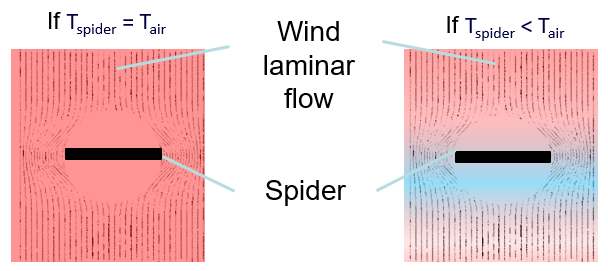}  
    \caption{Top : Subaru (left) and Keck (right) pupils. The spiders or segments fragment the pupil of the telescopes. Bottom : The low Wind Effect: a thermal effect. The temperature differential between the ambient air and the spiders ($T_{spider}<T_{air}$) induces radiative exchanges: the air gets cooler near the spider. This creates a refractive index gradient near the spiders that induces local delays of the wavefront. This phenomenon occurs especially when the wind speed is low (typically below $3$m/s), as the colder air is not blown away.}
    \label{fig-LWE}
\end{figure}

If the current XAO systems manage to correct for the atmospheric perturbations and therefore allow to obtain high angular resolution, limitations in achievable contrast still remain. First of all, the difference of light path between the "sensing" and the "science" paths makes so that several optical aberrations seen by the science detector are simply not seen by the Wavefront Sensor (WFS - see Fig.~\ref{fig-NCPA}). This generates static and/or quasi-static "speckles" in the science image that degrade the contrast. If the static speckles can be calibrated, the quasi-static ones remain a problem during observation. Second, the discontinuities in the pupil caused by the telescope spiders or segmented apertures (see Fig.~\ref{fig-LWE}-top), can also be the source of wavefront perturbations. Indeed the segmented telescopes need to be accurately co-phased (typically with a precision below $\lambda/40$~\cite{Harvey90}) for optimal use, or else the quality of the images is degraded. On the top of this, the spiders can be the cause of the Island Effect (IE)~\cite{milli2018low}, which generates differential aberration modes between the quadrants of the telescope (quite similar to aberrations corrected for the co-phasing of segmented apertures). These modes are even more triggered in the presence of low wind, generating the Low Wind Effect~\cite{sauvage2015low}: a thermal effect originating from radiative exchanges between ambient air and the spiders, inducing refractive index gradients near the spiders when the wind is typically below $3 m.s^{-1}$ (see Fig.~\ref{fig-LWE}-bottom). 
{}

Focal plane wavefront sensors have been largely studied to correct for NCPAs, Island Effect and segment co-phasing~\cite{Paxman-88,Redding-p-98,meimon2008phasing,korkiakoski2014fast,Vievard-a-17,vievard2019overview} since they can be operated directly on the science camera, avoiding NCPAs, and can sense discontinuities in the pupil's phase. As a competitive science instrument~\cite{Goebel2018,Currie2018,Currie2020,Lawson2020,norris2015vampires,jovanovic2017developing} and experimental platform, SCExAO~\cite{2015PASP..127..890J} (Subaru Coronagraphic Extreme Adaptive Optics), hosts and welcomes many new algorithms and concepts to improve the wavefront quality distributed to its different modules~\cite{huby2012first,walter2018mec,lagadec2018glint,rains2018development} , but also to prepare for the next generation of ELTs that will face similar challenges. The unique design of SCExAO allows to easily test focal plane wavefront sensors (among other WFS/C techniques, see paper No. $11448-78$ by \textit{O. Guyon}) using different cameras at different wavelengths with the internal calibration source, and gives an exclusive on-sky access for demonstration on an 8-meter class telescope.

\begin{figure}[!h]
    \centering
    \includegraphics[width=0.8\linewidth]{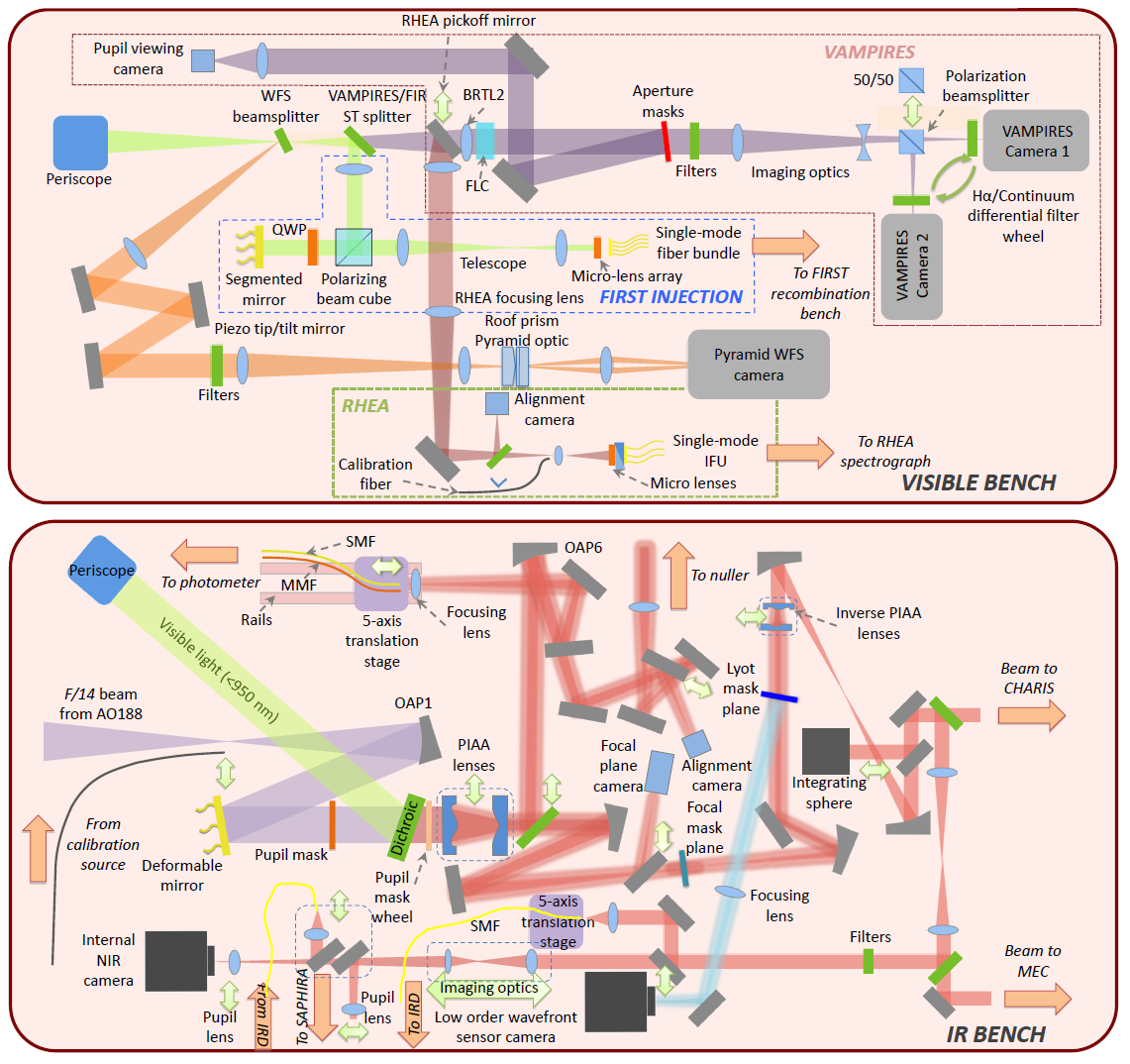}
    \caption{SCExAO IR and Visible bench.}
    \label{fig:scexao_visi}
\end{figure}{}

SCExAO gets a partially correct wavefront from the AO facility AO188~\cite{minowa2010performance}, and applies an XAO correction thanks to a pyramid wavefront sensor (PyWFS) operating in the visible~\cite{Lozi_2019} . As shown on Fig.~\ref{fig:scexao_visi}, SCExAO distributes the light to different modules and sensors scattered on two benches: a Visible and an Infra-Red bench. SCExAO experiences limitations in terms of wavefront quality or achievable contrast because of the previously mentioned NCPAs or IE, therefore it is the perfect place to test and validate different WFS/C techniques. In this paper, we focus on presenting a non-exhaustive list of six different focal plane wavefront sensors developed by local members of the SCExAO team or brought by collaborators. We succinctly present the principle of each technique, their implementation requirements on SCExAO and the latest obtained results.

%% file: 2-ZAP.tex
\section{The Zernike Asymetric Pupil wavefront sensor}
\subsection{Principle}\label{zap_ppe}
\begin{figure}[!h]
    \centering
    \includegraphics[width=0.5\linewidth]{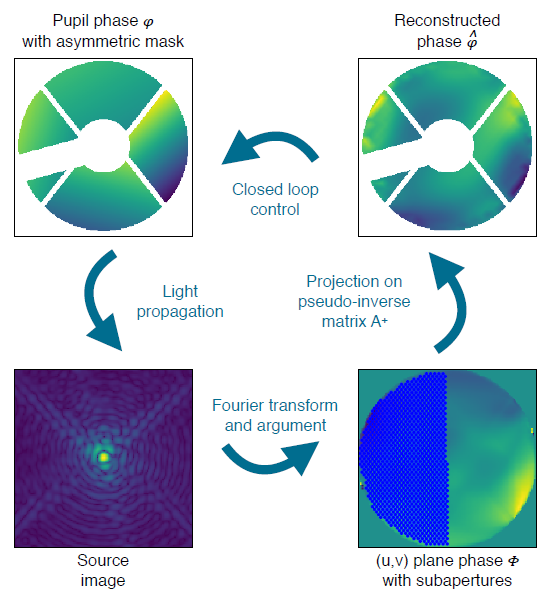}
    \caption{ZAP principle \cite{ndiaye_zap}}
    \label{fig:zap_principle}
\end{figure}

The first wavefront sensor we present is based on a Kernel phase analysis of the focal plane image of an unresolved source~\cite{martinache2013asymmetric} . In the small aberration regime ($\psi \ll 1$ rad), the relation between $\psi$, the phase in the pupil plane, and $\phi$ the phase in the Fourier space is linear:
\begin{equation}\label{ZAP_eq}
\phi = \mathbf{A}\psi,
\end{equation}
with $\mathbf{A}$ the transfer matrix between the two phases. This equation is easily inverted by computing $\mathbf{A^{\dag}}$ the pseudo-inverse of the rectangular matrix $\mathbf{A}$. The Singular Value Decomposition (SVD) of $\mathbf{A}$ gives: $\mathbf{A^{\dag}}=(\mathbf{A^{T}}\mathbf{A})^{-1}\mathbf{A^T}$. The inversion of Eq.~\ref{ZAP_eq} allows to compute the estimated phase aberration $\widehat{\psi}$ in the pupil plane from the phase $\phi$ measured in the Fourier space. However, $\widehat{\psi}$ suffers from a sign ambiguity. This degeneracy is lifted by introducing an asymmetry in the pupil plane, with a mask for example. Such a mask is shown Fig.~\ref{fig:zap_principle} top left. The process, shown on Fig.~\ref{fig:zap_principle} is therefore to compute the argument of the image Fourier transform and project the latest on the pseudo-inverse $\mathbf{A^{\dag}}$ to have an unambiguous estimation of the pupil phase $\widehat{\psi}$:
\begin{equation}\label{ZAP_res}
\widehat{\psi} = \mathbf{A}\phi.
\end{equation}
More details about the calibration of matrix \textbf{A} with Low Wind Effect modes from the instrument can be found in \textit{N'Diaye et al.}~\cite{ndiaye_zap}. 

{}

\subsection{Implementation on SCExAO}
\begin{figure}[!h]
    \centering
    \includegraphics[width=0.8\linewidth]{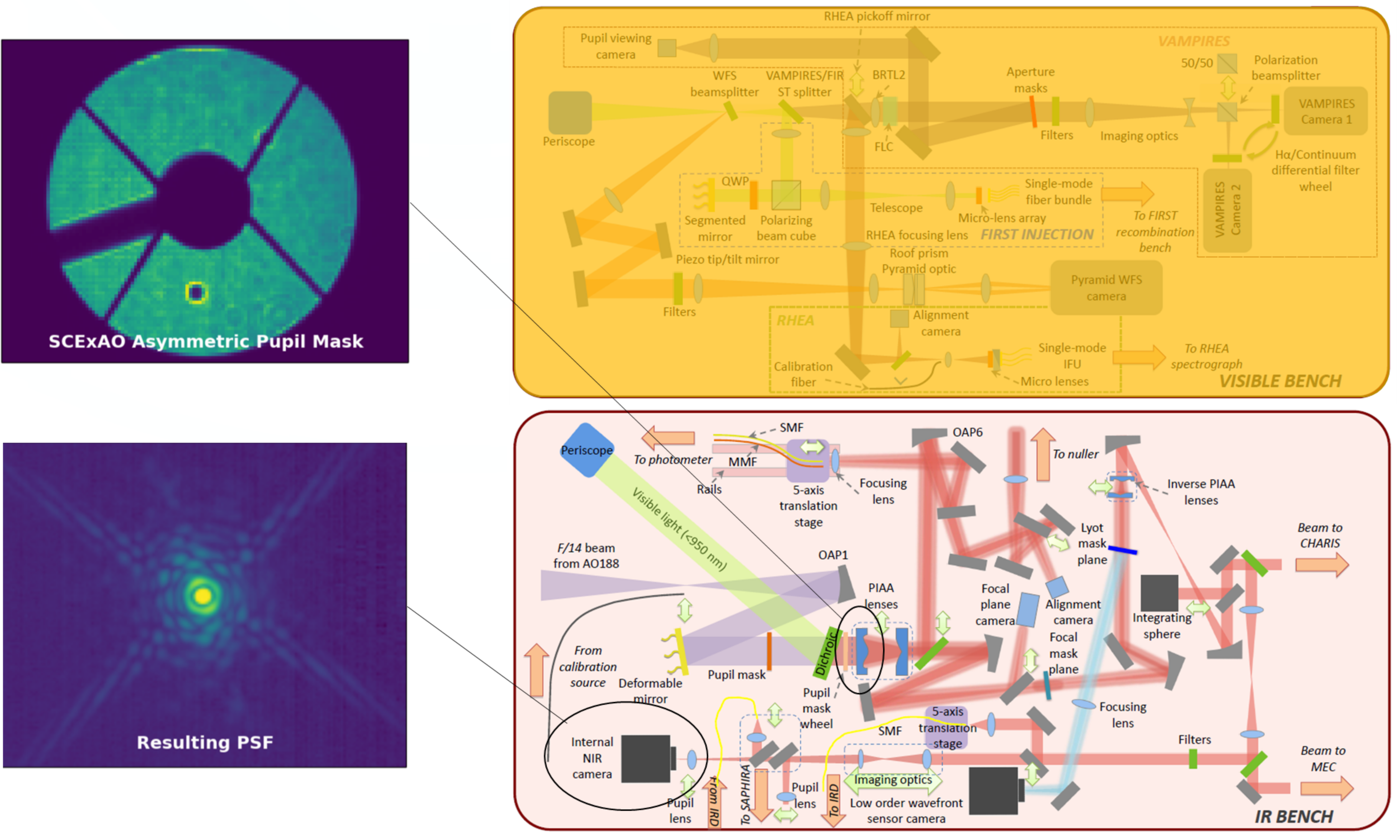}
    \caption{ZAP implementation in SCExAO: the asymmetric pupil mask is placed in a pupil plane filter wheel upstream the internal NIR camera. Images can then be recorded on the latest camera.}
    \label{fig:zap_implem}
\end{figure}

The requirements for the integration of the Zernike Asymetric Pupil wavefront sensor are :
\begin{itemize}
    \item a pupil plane mask,
    \item a focal plane image.
\end{itemize}
As we can see Fig.~\ref{fig:zap_implem} on the bottom right, SCExAO IR bench has a pupil mask wheel in the light path upstream the CRED 2 internal Near IR camera. The implementation is then straightforward and one of the pupil mask wheel slot contains the asymmetric mask shown Fig.~\ref{fig:zap_implem} top left. Images are then acquired with the internal NIR camera.

{}

\subsection{Results: on-sky}



\begin{figure}[!h]\centering
	\includegraphics[width=0.3\linewidth]{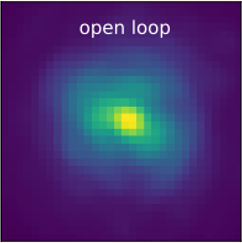}  \includegraphics[width=0.3\linewidth]{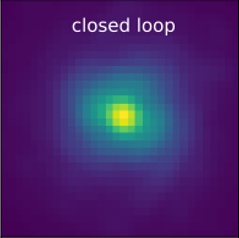}
\caption{Closed loop operation of ZAP on-sky~\cite{ndiaye_zap}. Focal image (left) in presence of LWE shows bright secondary lobes. Once the loop is closed with ZAP, the quality of the wavefront is improved, the bright lobes disapear (right image). A relative Strehl ratio of 37\% was measured between the two images.}
\label{fig:closeloop-sky-ZAP}
\end{figure}

The algorithm was tested on-sky~\cite{ndiaye_zap} in a closed-loop operation, in presence of LWE. Fig.~\ref{fig:closeloop-sky-ZAP} shows on the left the open loop, the PSF in presence of LWE acquired with the SCExAO/VAMPIRES module, in the visible (@$750$nm). We can see bright secondary lobes, sign of LWE. Image on the right shows the state of the PSF after closing the loop. An estimation of the Strehl ratio by measuring the relative intensity in those two images gives an improvement of $37\%$ before and after loop-closure.

%% file: 4-LAPD.tex
\section{The Linearized Analytic Phase Diversity}
\subsection{Principle}\label{lapd_ppe}
\begin{figure}[!h]
    \centering
    \includegraphics[width=0.5\linewidth]{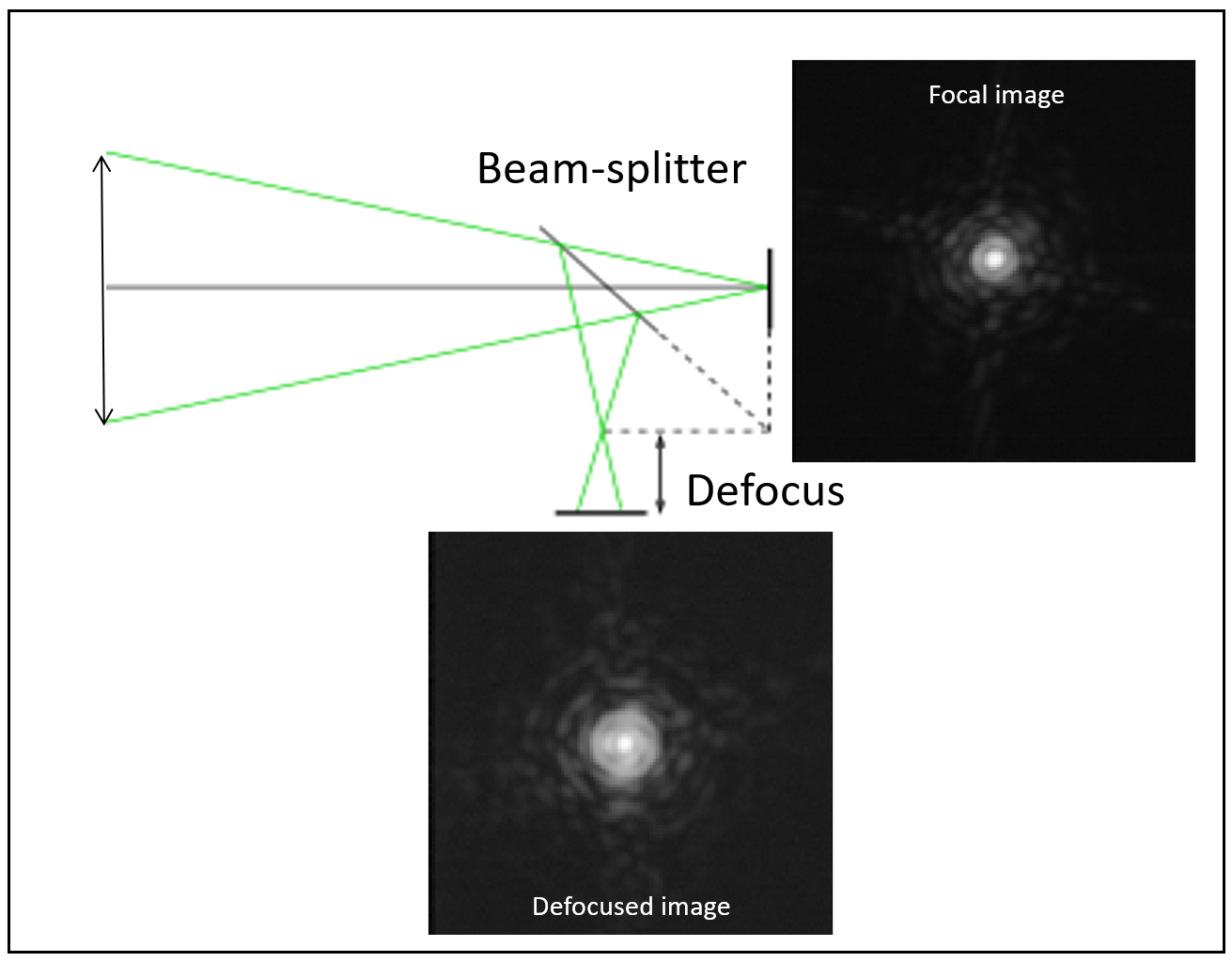}
    \caption{Phase diversity principle: simultaneous acquisition of focal and defocused images.}
    \label{fig:lapd_ppe}
\end{figure}

As its name indicates, the Linearized Analytic Phase Diversity (LAPD) algorithm is based on the phase diversity technique~\cite{Gonsalves-a-82,Mugnier-l-06a} , in the case of small aberrations. Phase diversity technique consists in the acquisition of two images: a focal plane image and an image with a known aberration (here, a defocus - see Fig.~\ref{fig:lapd_ppe}).

{}

The acquisition of the second image allows to lift the degeneracy of the phase estimation based on the sole focal plane image (see section~\ref{zap_ppe}). The two images are used to compute a criteria that is defined as the distance between the images and a model of the telescope images. The aim is then to minimize this criteria. Since the link between the aberrations and the images is non-linear, it usually requires time-consuming, high computational cost iterative minimization algorithms.

LAPD, originally developed in the context of segmented telescope fine phasing~\cite{vievard2020cophasing} , offers performance similar to the phase diversity technique but with a lower computing cost. Under the assumption of small aberrations, we can compute a $1^{st}$ order Taylor expansion of the PSF, leading to an analytical form of the phase diversity problem (more details provided in \textit{Mocoeur et al.}~\cite{Mocoeur-a-09b}). Simply doing this only offers a limited capture range of the algorithm. The idea behind LAPD is to exploit this linearization and operate it iteratively, similarly to the Newton-Raphson method~\cite{ypma1995historical} .

We can then consider the Subaru telescope as a 4 sub-aperture telescope. Each of the IE modes can be decomposed as piston/tip/tilt in each quadrant, acting like a multi-aperture mirror. LAPD is also able to estimate higher order aberrations on the full pupil.

\subsection{Implementation on SCExAO}
The requirements for the integration of the LAPD wavefront sensor are :
\begin{itemize}
    \item a focal plane image,
    \item a defocused plane image.
\end{itemize}

LAPD is implemented on the VAMPIRES cameras. In its nominal configuration, VAMPIRES operates with two focal images. We added a lens to obtain a defocus on one of the two cameras, allowing to simultaneously acquire a focal plane and a defocused plane.

\begin{figure}[!h]
    \centering
    \includegraphics[width=0.75\linewidth]{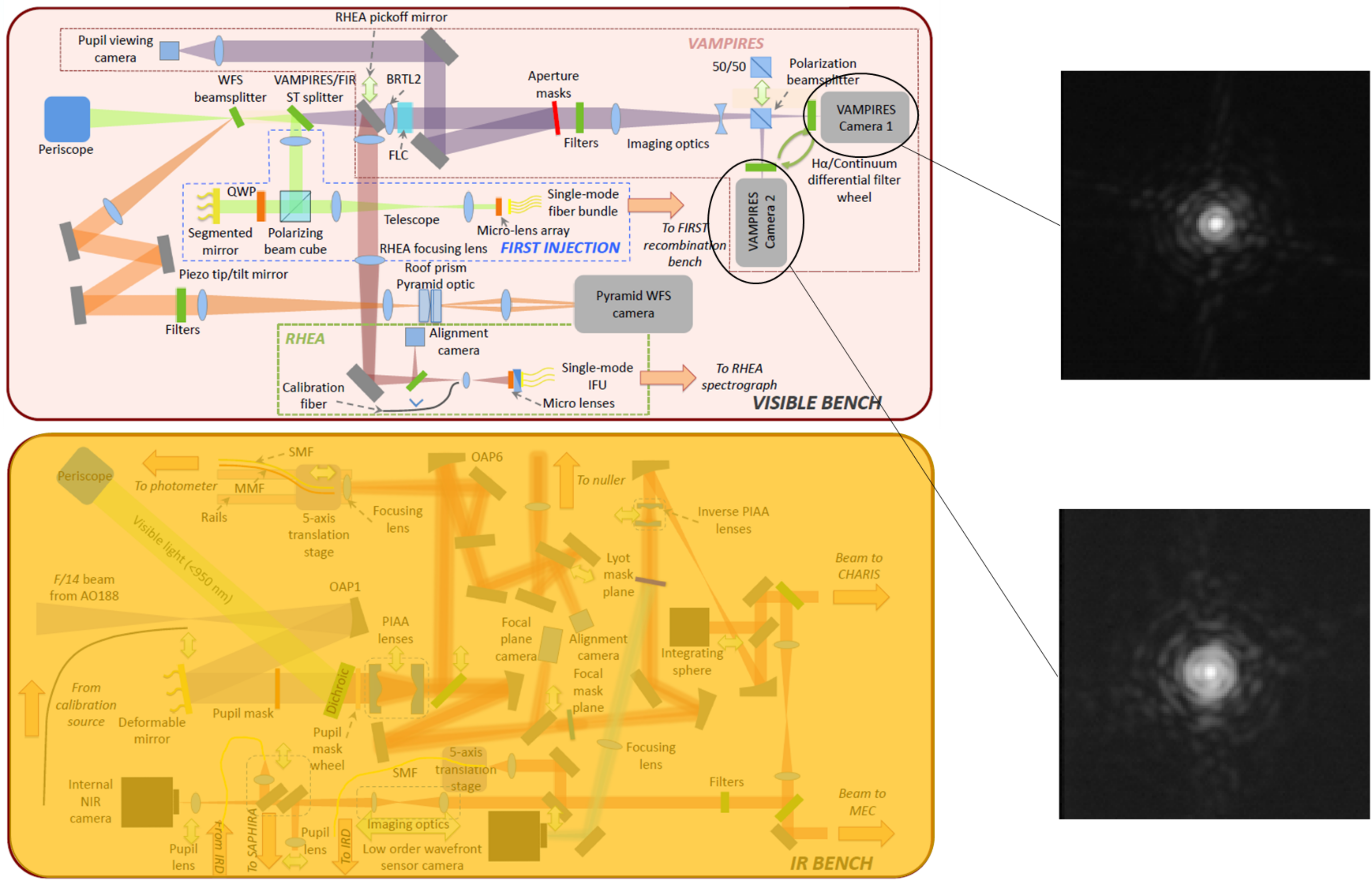}
    \caption{LAPD implementation in SCExAO: Two images can simultaneously be acquired with two cameras in the VAMPIRES module. Both cameras make focal plane cameras, but one sits on a translation stage allowing to introduce the desired phase diversity (defocus).}
    \label{fig:lapd_implem}
\end{figure}{}

\subsection{Results: in-lab}

\begin{figure}[!h]\centering
	\includegraphics[width=\linewidth]{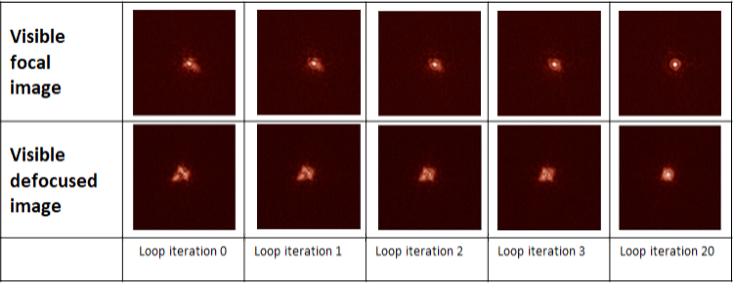}
\caption{Closed-loop sequence with LAPD to correct for IE perturbations on SCExAO internal source.}
\label{fig:estim-simu-lapd}
\end{figure}

We integrated LAPD and tested it using the SCExAO internal source. We applied an IE-like static phase map on the deformable mirror (with Piston/Tip/Tilt aberrations in each quadrants), and ran LAPD in a closed-loop sequence. The RMS value used for our test was $0.5$ rad RMS. We show the sequence on Fig.~\ref{fig:estim-simu-lapd}. The first and second rows of Fig.~\ref{fig:estim-simu-lapd} respectively show the focal and defocused images during the sequence. We can see at loop iteration 0 that the PSF presents asymmetries that are typical from IE aberrations. Along the loop sequence, performed with a gain set to $0.3$, we can see the ability of LAPD to converge to a state where the PSF (loop interation 20) is symmetric, sign that the IE aberrations were corrected. LAPD will be tested on-sky during the next Engineering run campaign. 

%% file: 5-MonoPD.tex
\section{Single image Phase Diversity}

\subsection{Principle}\label{monoPD_ppe}

As discussed in Section~\ref{lapd_ppe}, Phase Diversity relies on the acquisition of two images with a differential known aberration (\textit{e.g.} a defocus). However, it not always easy to simultaneously acquire them. In the context of characterizing the Low Wind Effect on the SPHERE instrument, \textit{M. Lamb}~\cite{Lamb-ao4elt5} showed that it was possible to apply Phase Diversity with a single image. To perform this, two conditions must be considered: the observed object has to be unresolved by the telescope and the defocus of the diversity image must be large enough.

\subsection{Implementation on SCExAO}

There is only one requirement for the integration of the mono-plan Phase Diversity:
\begin{itemize}
    \item a defocused plane image.
\end{itemize}

We use the NIR camera for the integration of the algorithm. The camera was translated on a rail to acquire images with $1\mu$m defocus. 

\begin{figure}[!h]
    \centering
    \includegraphics[width=0.8\linewidth]{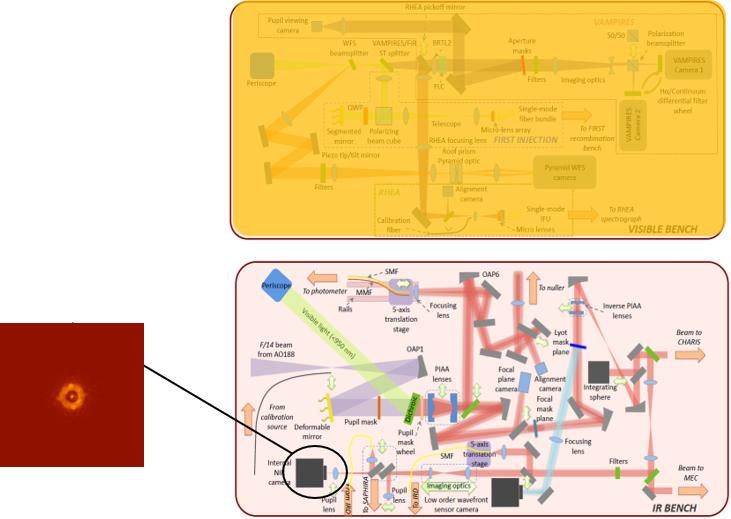}
    \caption{Mono-plan Phase Diversity implementation in SCExAO only requires a defocused image from the NIR camera.}
    \label{fig:ff_implem}
\end{figure}{}

\subsection{Results: in-lab}

Similarly to the in-lab validation of LAPD, we applied static aberrations on the SCExAO deformable mirror to test the algorithm. Because the algorithm is not yet implemented on SCExAO, and because our collaborator is not physically present in Hawaii, the way we proceeded was to acquire several images perturbed with IE modes, and we emailed them along with the deformable mirror telemetry. This allowed to proceed with open-loop tests before a full integration of the algorithm on the bench.

\begin{figure}[!h]
    \centering
    \includegraphics[width=\linewidth]{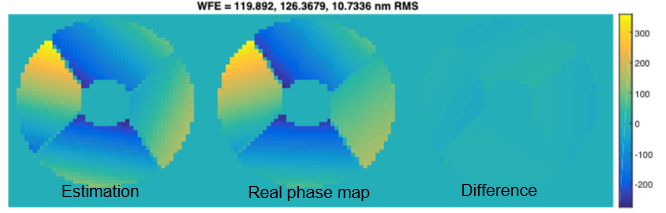}
    \caption{Left : Estimation map delivered by the single image phase diversity algorithm. Middle : Applied static phase map on SCExAO deformable mirror. Right : Residual aberrations after open-loop estimation.}
    \label{fig:monoPD}
\end{figure}

One of the open-loop test is presented on Fig.~\ref{fig:monoPD}. The central image showed corresponds to the applied phase map on the deformable mirror, where we can distinguish the differential piston, tip and tilt aberrations in the four different quadrants of the pupil. The amplitude of the perturbations was $126.4$nm RMS at $1\mu$m (or about $0.6$~rad RMS). The estimated phase map by the single image phase diversity algorithm is shown on the left. We can visually assess that it resembles to the applied phase map. A look at the difference on the right, or residual map, confirms that the estimation matches well the reality. Indeed the residual phase amplitude is about $11$nm RMS at $1\mu$m, or about 0.06~rad RMS.

{}

This confrms the ability of the single image phase diversity to estimate IE-induced aberration on SCExAO. The next step is to test open-loop estimations on on-sky images acquired in the presence of IE perturbations.

%% file: 6-FandF.tex
\section{Fast \& Furious (sequential phase diversity)}

\subsection{Principle}

We present now a sequential phase diversity wavefront sensor: Fast and Furious~\cite{keller2012extremely,korkiakoski2014fast} (F\&F). This algorithm was already studied in simulation and the lab for a SPHERE implementation to control the LWE~\cite{wilby2016fast,wilby2018laboratory}. Instead of spatial phase diversity, as discussed in section~\ref{lapd_ppe}, this algorithm relies on a temporal phase diversity. The algorithm reconstructs the wavefront by splitting a single PSF image in its even and odd components. The odd component directly yields the odd wavefront error by simple algebra. For the even component one image is not sufficient to reconstruct the even wavefront error as the reconstruction suffers from a sign ambiguity. This degeneracy is lifted by the use a second PSF image with a known phase diversity, which is provided by the previous iteration and DM command of the F\&F loop. This has the major advantage that the wavefront sensor can be operated simultaneously with science observations as it continuously improves the wavefront. The algorithm is computationally very efficient as it requires only one Fourier transform and simple algebra per iteration.   

\subsection{Implementation on SCExAO}
\begin{figure}[!h]
    \centering
    \includegraphics[width=0.8\linewidth]{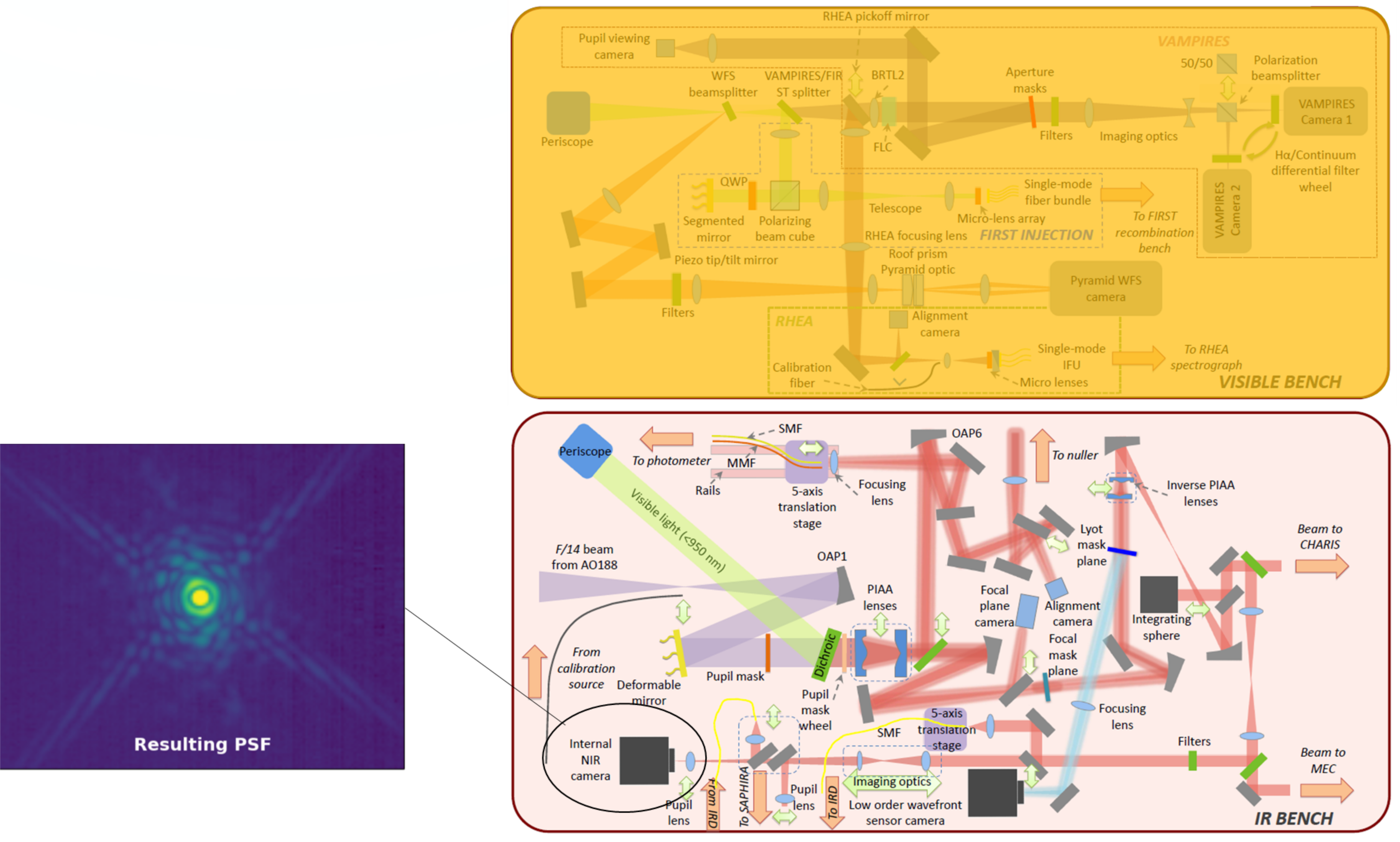}
    \caption{F\&F implementation in SCExAO only requires an image from the NIR camera and the DM stream.}
    \label{fig:ff_implem}
\end{figure}

There are two requirements for the integration of the F\&F WFS in SCExAO:
\begin{itemize}
    \item a focal plane image,
    \item an access to the DM.
\end{itemize}
For these simple requirements, we can just use the CRED 2 internal Near IR camera (see Fig.~\ref{fig:ff_implem}) and the DM stream. We can close the loop on the DM with F\&F while it is fed by the NIR camera images.

{}

\subsection{Results: on-sky}

\begin{figure}[!b]\centering
\includegraphics[width=\linewidth]{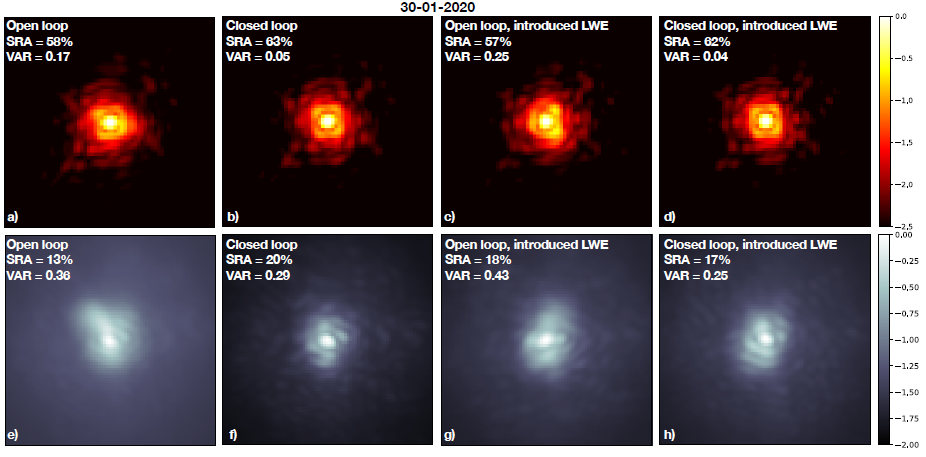}
\caption{Demonstration of F\&F on-sky to correct for the Low Wind Effect~\cite{refId0}. Top row - NIR images, used for F\&F operation. Bottom row - Monitoring of visible PSF on the VAMPIRES focal plane sensor.}
\label{fig:closeloop-onsky-FF}
\end{figure}

On-sky tests were conducted on the Rigel target ($m_H=0.2$) in January 2020 (see Fig.~\ref{fig:closeloop-onsky-FF})~\cite{refId0}. If F\&F was operated on the NIR, we also monitored the visible focal image thanks to the VAMPIRES camera. \\

We can see on (a) the on-sky PSF without correction. The asymmetries that appear in the first Airy ring are typically originated from the LWE. A first close-loop test performed shows that the PSF symmetry was restored~(b). In order to push the test a little further, we applied static LWE modes on the DM (similarely to the tests for LAPD and the single image phase diversity, described above). (c) shows the resulting PSF, strongly affected by the island modes. Again, a close-loop with F\&F was efficient to correct for these introduced LWE modes (d). Both times, the estimated gain in Strehl ratio was of 5\%. A visual assessment of the visible images also shows the correction of the island modes. The success of these test allow to confirm the ability of F\&F to correct for the LWE, but it can also be very useful for NCPAs. Current efforts are focused on making the algorithm faster, and deploying it on all the different science detectors on SCExAO.

%% file: 7-NN.tex
\section{PSF reconstruction from PyWFS using Neural Network}
\subsection{Principle}
This method developed by \textit{Norris and Wong, 2019}, aims at training a deep Neural Network (NN) with synchronized in time PyWFS telemetry and focal plane images. Even though the training is slow and requires a large amount of data, its application can be done in real-time on modern GPUs equipped with tensor cores.

\subsection{Implementation on SCExAO}
The requirements for the integration of the NN are :
\begin{itemize}
    \item a focal plane image,
    \item the PyWFS telemetry.
\end{itemize}

\begin{figure}[!h]
    \centering
    \includegraphics[width=0.8\linewidth]{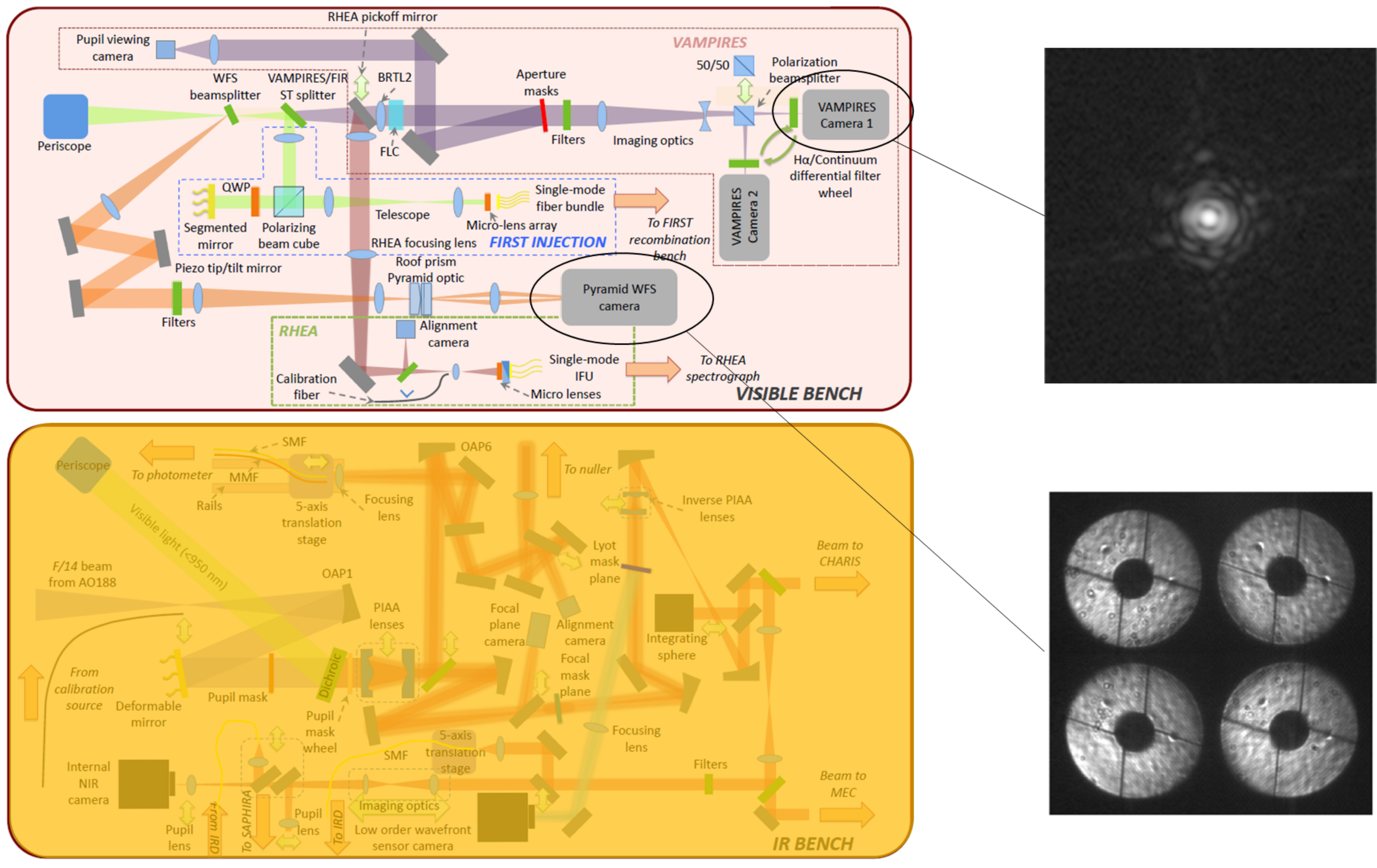}
    \caption{NN implementation in SCExAO: training of the NN needs the telemetry of the PyWFS and the focal plane image from one of VAMPIRES camera.}
    \label{fig:nn_implem}
\end{figure}{}

We can see Fig.~\ref{fig:nn_implem} on the left that the NN is fed by one of VAMPIRES visible camera and the existing telemetry of the Pyramid. No extra hardware had to be added to implement the NN.

\subsection{Results: on-sky}
\begin{figure}[!h]
    \centering
    \includegraphics[width=0.9\linewidth]{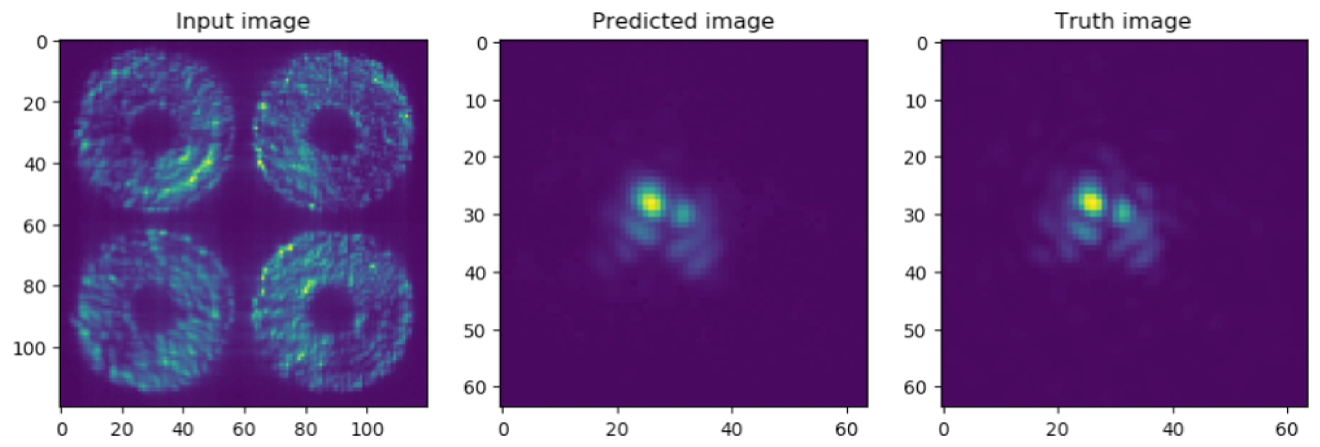}
    \caption{Neural Network on-sky results. Left: Input image for the NN, the PyWFS telemetry. Middle: Predicted image. Right: Real image. Courtesy: \textit{Barnaby Norris, University of Sydney}.}
    \label{fig:nn_onsky}
\end{figure}

Fig.~\ref{fig:nn_onsky} shows the on-sky demonstration of PSF prediction from the PyWFS telemetry. The left image is the PyWFS telemetry, the input for the NN, the central image is the predicted image by the NN, and the image on the right is the true PSF image. The two images, which are parts of a movie of several seconds, are visually very similar. It is to be noted that the visible PSF reconstruction is highly non-linear and particularly sensitive to small wavefront errors.

{}

%% file: 8-DRWHO.tex
\section{Dr WHO: Direct Reinforcement Wavefront Heuristic Optimization}

\subsection{Principle}\label{drwho_ppe}
The NCPAs are a considerable issue in direct imaging of exoplanets as it is the main limiting source of error in the PSF. Furthermore, in a system with a PyWFS, this problem is intrinsically linked with it. The key issue is to understand what is the \textit{true} reference of the PWFS, that we will call the \textit{Ideal reference}. In other word, we need to understand what gives a flat wavefront, corrected from all static and quasi-static aberrations and mainly the NCPA. 
The way the reference of the PWFS is currently set, is with an internal light source inside the instrument, before the observations. However, in reality, the \textit{ideal} PWFS reference is constantly evolving\footnote{because of the quasi-static aberrations and the chromaticity of the wavefront}, and is different on sky than with the internal light source. Therefore, a continuous way to measure the \textit{ideal} reference is essential. The Dr WHO algorithm \cite{drwhoskaf} aims to address this issue by continuously updating the reference of the PWFS through a lucky imaging of the focal plane fast camera. Every 30 seconds, the algorithm looks at the best 10\% focal plane images, take the corresponding WFS images, average them in one image and apply this as a new reference to the PyWFS, with an integrator filter. Thus, the algorithm is continuously rewarded for better focal plane images, the converges towards a better correction of the NCPA.

\subsection{Implementation on SCExAO}
The requirements for the integration of the Dr WHO are:
\begin{itemize}
    \item focal plane images,
    \item PWFS images.
\end{itemize}

\begin{figure}[!h]
    \centering
    \includegraphics[width=0.75\linewidth]{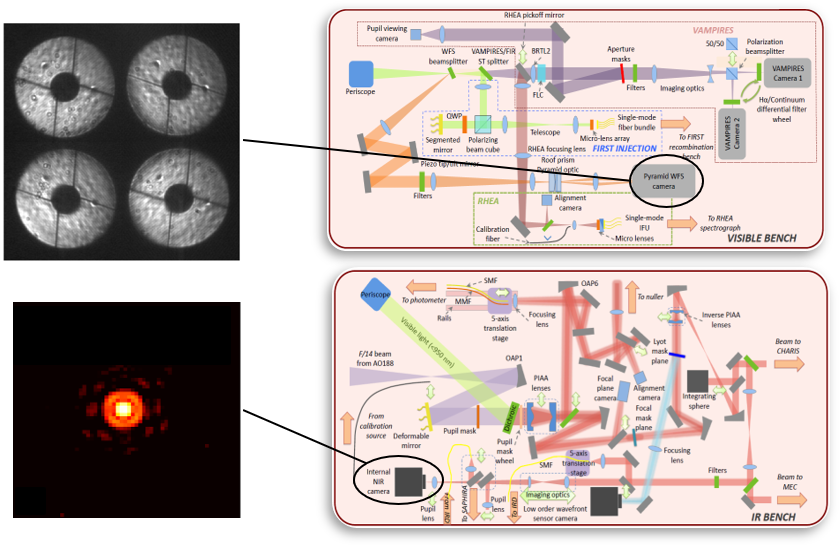}
    \caption{DRWHO implementation in SCExAO: the algorithm requires telemetry of the PyWFS and the NIR fast focal plane image.}
    \label{fig:drwhobench}
\end{figure}

Any focal plane fast camera can be used, especially we have been  using the the VAMPIRES cameras for performing testing the algorithm. Figure \ref{fig:drwhobench} emphasizes the data streams required. 

{}

\subsection{Results: on-sky and simulation}

\begin{figure}[!h]
    \centering
    \includegraphics[width=0.7\linewidth]{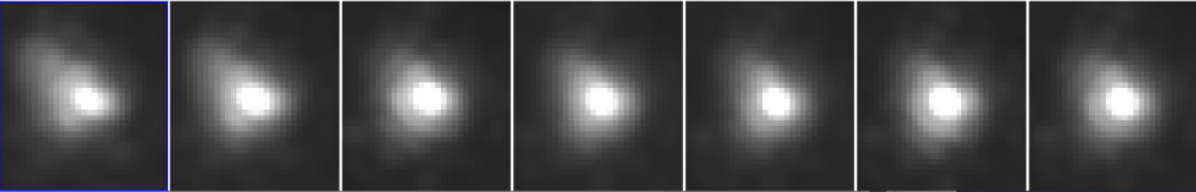}
    \caption{NCPA correction using VAMPIRES focal plane image feedback with Dr WHO.}
    \label{fig:drwhovampires}
\end{figure}

Dr WHO was first implemented very simplistically on sky in order to briefly test the relevance of the algorithm. Figure \ref{fig:drwhovampires} presents the evolution of the PSF throughout a 20 min sequence of the algorithm. \\

{}

\begin{figure}[!h]
    \centering
    \includegraphics[width=0.5\linewidth]{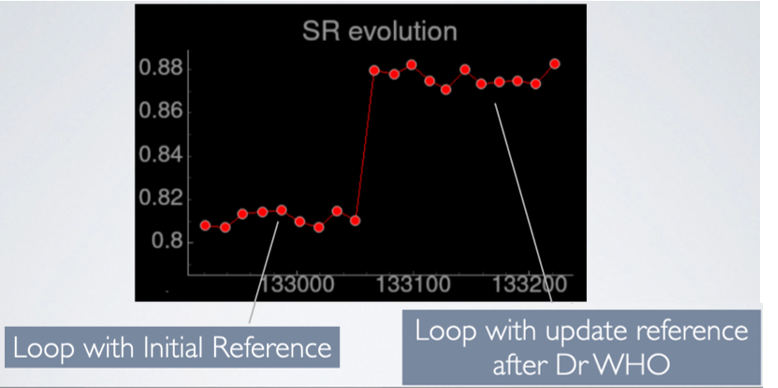}
    \caption{Compass simulation results of Dr WHO: the system without NCPA has a SR of 92\%, we added NCPA to affect the SR of abour 10\%. After a few iterations of Dr WHO, the new reference gives a sharp increase of the SR.}
    \label{fig:drwhocompass}
\end{figure}

We then performed more thorough tests with the Compass simulation software \cite{Compass} to better quantify the NCPA correction. This was done by first simulating SCExAO. We performed several tests by adding NCPA and running Dr WHO while the AO loop is running, Figure \ref{fig:drwhocompass} presents the difference of the Strehl ratio (SR) before and after the training of Dr WHO, in simulation. Further tests on the internal source of SCExAO are ongoing. 
{}

%% file: 9-Conclusion.tex
\section{Conclusion}
\begin{figure}[!h]\centering
\includegraphics[width=\linewidth]{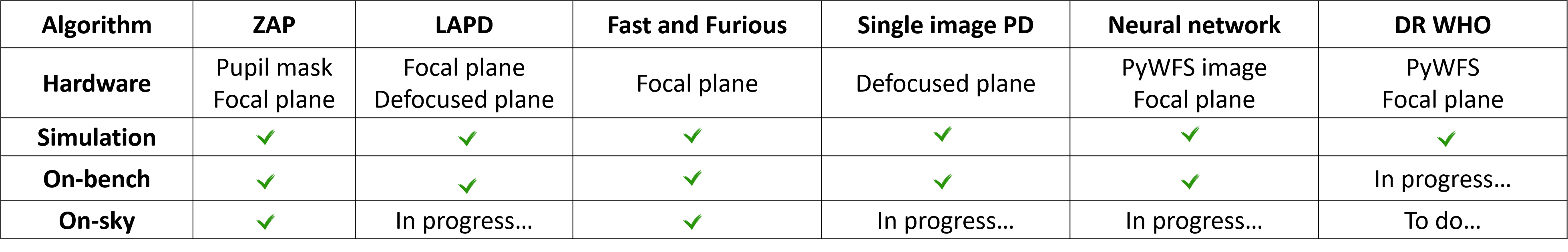}
\caption{Table of comparison between the different focal plane algorithms on SCExAO: Hardware requirements and progress in the integration on SCExAO.}
\label{fig:ccl}
\end{figure}

We presented six focal plane wavefront sensors implemented on SCExAO for the correction of NCPAs or Island Effect. The appealing aspect of these WFS is that their integration in a system is very easy. Fig.~\ref{fig:ccl} shows a table that summarizes the hardware requirements for each of the different algorithms, and their integration/validation progress on SCExAO. We can see that two algorithms have already been validated on-sky, three algorithms have been validated on the bench and will hopefully be soon demonstrated on-sky. One algorithm is currently being integrated on SCExAO, but has already shown promising simulation results. 

The modularity of SCExAO and the easy access to the hardware makes it a key platform to test and validate new concepts and algorithms. We showed here only a small portion of the WFS/C research that is being done on SCExAO. All this work provides unique opportunities to start tackling identified and upcoming WFSC/HCI issues for the future extremely large telescopes.